\documentclass{mem}
\usepackage{natbib}\usepackage{txfonts}\usepackage{balance}
\usepackage{graphicx}
\usepackage[a4paper]{hyperref}
\idline{75}{282}
\begin{document}
\def\teff{$T\rm_{eff }$}
\def\kms{$\mathrm {km s}^{-1}$}
\def\cntrt{counts cm$^{-2}$s$^{-1}$keV$^{-1}$}

\title{
Status of the Simbol-X detector background simulation activities
}

\subtitle{}

\author{
R. \,Chipaux\inst{1}
\and U. \,Briel\inst{2}
\and A. \,Bulgarelli\inst{3}
\and L. \,Foschini\inst{3}
\and E. \,Kendziorra\inst{4}
\and C. \,Klose\inst{5}
\and M.~\,Kuster\inst{5}
\and P. \,Laurent\inst{1}
\and C. \,Tenzer\inst{4}
}

\offprints{R. Chipaux, \email{remi.chipaux@cea.fr}}

\institute{
CEA/DSM/Dapnia, CEA-Saclay, 91191 Gif sur Yvette Cedex, France
\and
Max-Planck-Institut f\"ur extraterrestrische Physik, Giessenbachstr.,
85748 Garching, Germany
\and
Istituto di Astrofisica Spaziale e Fisica Cosmica, Via Gobetti 101,
40129 Bologna, Italy
\and
Institut f\"ur Astronomie und Astrophysik, Universit\"at T\"ubingen, Sand 1,
72076 T\"ubingen, Germany
\and
Institut f\"ur Kernphysik, TU Darmstadt, Schlossgartenstr. 9, 64289 Darmstadt,
Germany
}

\authorrunning{Chipaux et al.}

\titlerunning{Simbol-X background simulation activities}

\abstract{
Using the Geant4 toolkit, a Monte-Carlo code to simulate the detector background
of the Simbol-X focal plane instrument has been developed with the aim to
optimize the design of the instrument.
Structural design models of the mirror and detector satellites have been
built and used as baseline for our simulations, to evaluate the different
background contributions that must be taken into account to determine
the sensitivity of the Simbol-X detectors.
We work towards a simulation based background and mass model which can be
used before and during the mission.

For different material compositions, material thicknesses, locations etc.
the response of the instrument to the diffuse cosmic hard X-ray background and
to the cosmic proton induced background have been calculated.
As a result we present estimates of the background count rate expected in the
low and high energy detector, and anti-coincidence rates.
The effect of induced radioactivity in the detector and shielding materials and
soft proton scattering in the mirror shells are also under study.
}
\maketitle{}

\section{Introduction}
The focal plane detector of Simbol-X is made of two layers:
a monolithic DEPFET Silicon detector is placed on top of a pixelized CdTe or
CdZnTe detector.
To achieve the scientific goals of the mission, an ambiguously low background
level of below 10$^{-4}$ \cntrt ~is required.
This is far below the measured background of X-ray CCD cameras currently flown
on missions like XMM-Newton and Chandra.
The Monte-Carlo toolkit Geant4 is used by the Simbol-X background simulation
group to optimize the design of the detector assembly in terms of low
background.
The background of the Simbol-X focal plane detector mainly stems from four
components:
the diffuse cosmic photon background, background induced by prompt interactions
of high energy protons, a delayed component from induced radioactivity and soft
protons funnelled by the telescope onto the detector.
All four components are studied by our group in detail.

\section{Simulations and models}

\subsection{Geant4}
Geant4 is a toolkit for the simulation of the passage of photons and particles
through matter.
Its areas of application include high energy, nuclear and accelerator physics,
as well as studies in medical and space science \citep{Agos03,Alli06}.

In a preliminary study the model included only electromagnetic
interactions~\citep{Tenz06}.
In the work reported here the release 8.2.p01 of Geant4 was used and hadronic
interactions were added.
In addition the radioactive decay module of Geant4 was used for
section~\ref{radioactivity},
and the scattering of low energy protons module for section~\ref{optics}.

\subsection{Simulation Models}

\subsubsection{Detectors}
\begin{figure*}[]
\resizebox{\hsize}{!}{\includegraphics[clip=true]{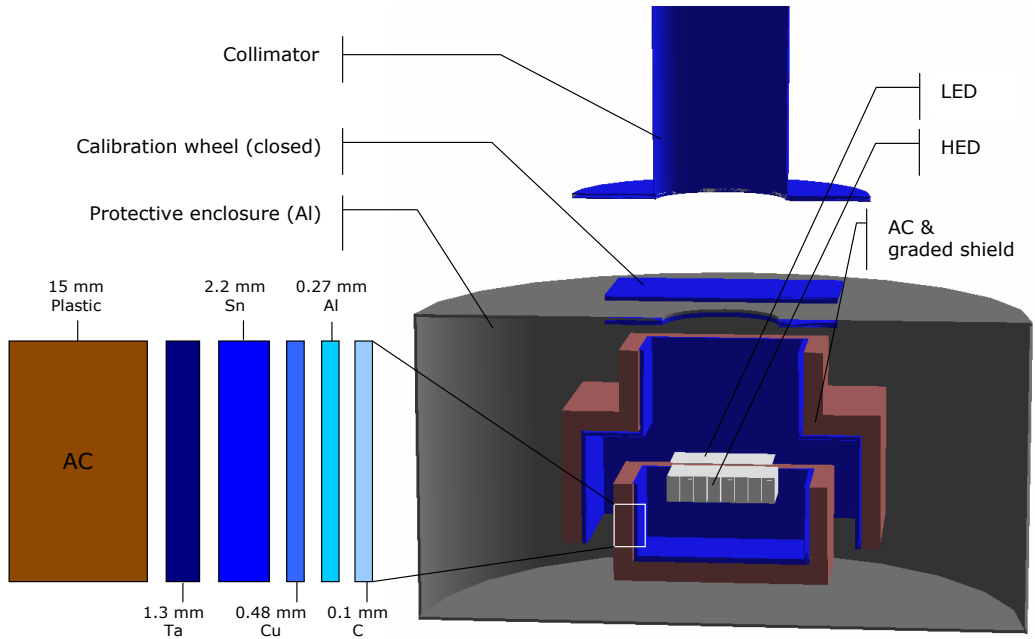}}
\caption{
\footnotesize
Mass model of the detectors, used for Geant4 Monte Carlo simulations.
}
\label {modelDet}
\end{figure*}

A model, shown in Fig.~\ref{modelDet}, of the current baseline configuration
of the detector housing and the detectors has been built using Geant4 tools.
This simplified geometry omits cables and some structures and details,
but otherwise accurately reproduces the mechanical design as of January 2007.
The central elements of the geometric model are the two detectors.
The low-energy detector (LED) is represented by a slab of silicon, dimensions
$80\times 80\times 0.45 $~mm$^3$, without segmentation.
The high-energy detector (HED) is composed of $8\times 8$ modules of cadmium
zinc telluride (Cd$_{0.9}$Zn$_{0.1}$Te), $10\times 10\times 2$~mm$^3$,
separated by 0.625~mm gaps.
The segmentation of the HED is not yet used, the results presented refer
to the sum of events in the 64 modules.
The electronics associated to each module is represented by a
$10\times 10\times 17.5$~mm$^3$ box in gold.

The detectors are surrounded by an anti-coincidence shield (AC), which consists
of plastic scintillator slabs, 15~mm thick, divided in top, lateral and bottom
parts to allow space for read-out connections.
A graded shield is foreseen inside AC to reduce the incoming photon flux.
It is designed to absorb all photons below 200~keV,
leaving X-ray fluorescence below 0.3~keV and therefore below the detection
limit of the LED (see Fig.~\ref{modelDet} for details on the composition and
thicknesses).
An aluminum structure encloses and stabilizes the active parts of the camera.

A 1.6~m-long collimator sits above the aperture of the detector box to
prevent photons coming from outside the field of view from hitting the
detectors.
It consists of the same materials as the graded shield.
Its thickness, however, is decreasing with the distance from the detector to
save weight and to maintain a constant effective thickness with respect to the
incident angle of the incoming radiation.

A ``calibration wheel'', with the same structure as the graded shield,
is placed between collimator and detector box.
It allows us to either open or close the field of view.
As a general rule, the cosmic photon-induced background is calculated with the
field of view closed, the proton-induced one with the field of view open.
Anyway, the wheel is practically transparent to cosmic protons.
In the setup used here the spacecraft below the detector was not taken into
account.

This model allows to estimate the effects of a change of materials,
thicknesses or positioning on the detector background and thus helps to
optimize the design of the detector system.

\subsubsection{Particle spectra and fluxes}

The isotropic cosmic flux impinging the spacecraft is simulated by emitting
particles from the inner surface of a sphere of radius larger
than the spacecraft overall dimensions. 
To save computing time, the direction of emission is restricted to a cone
containing the spacecraft.

\begin{figure}[]
\resizebox{\hsize}{!}{\includegraphics[clip=true]{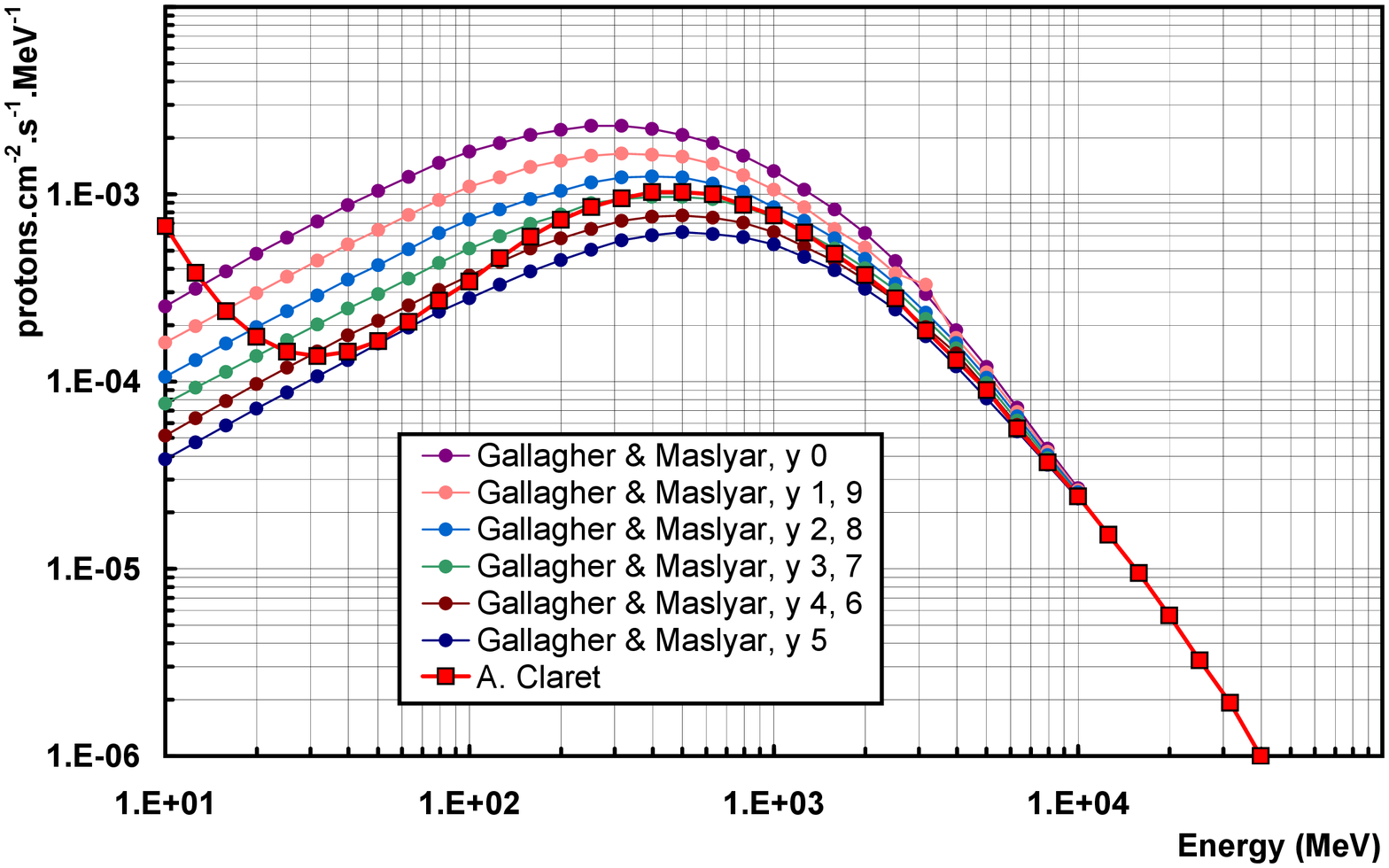}}

\caption{
\footnotesize
Proton spectrum calculated by \citet{Clar06} (squares), as used in the present
simulation.
For comparison, the spectra published by \citet*{Gall76} for different periods
of the solar cycle are also shown.
}
\label{spectreProtons}
\end{figure}

In this study, we restrict ourselves to cosmic photon and proton fluxes,
neglecting electrons and ions.
The photon flux is supposed to follow the intensity and spectrum published by
\citet{Grub99} for the diffuse cosmic hard X-rays.
We extrapolate the analytic formula given by \citeauthor{Grub99} down to 1~keV
and take an upper limit equal to 100~MeV.
In this range the integral of the X-ray flux is equal to
197.3~photons~cm$^{-2}$s$^{-1}$ over $4\pi$.
For the protons we use the spectrum computed by \citet{Clar06}, at the expected
launch date (near solar maximum), between 10~MeV and 100~GeV.
In this range the integral of the proton flux is estimated
to 2.31~protons~cm$^{-2}$s$^{-1}$ over $4\pi$.
As shown in Fig.~\ref{spectreProtons}, this spectrum is comparable to the
spectra published by \citet{Gall76} for different periods of the solar cycle.

\subsubsection{Optics}
Concerning the mirror, the model is derived at the present state from the one
developed for XMM \citep{Nart02}.
The optics is composed of 58~shells at 20~m from the detectors.


\section{Results}

%
%

\subsection{Detectors prompt background}

As reported in table~\ref{tabcr}, in the current setup the background count
rates are dominated by the cosmic protons.
The cosmic photons induce a count rate of about $10^{-4}$ \cntrt~in LED and
$3~10^{-4}$ in HED.
As expected, the AC is not efficient in that case.
The cosmic protons induce a much higher background, in the range of about
$3~10^{-3}$ \cntrt~in both detectors.
The AC plays here its full role and allows reduction of the effective proton
background to the range of $10^{-4}$ \cntrt.

\begin{table}
\caption{Count rates (in $10^{-5}$ \cntrt) in LED and HED due to cosmic photons
and protons for various configurations of AC:
a) basic setup with AC made of plastic scintillator;
b) AC made of NaI;
c) same as a) but protective enclosure in Ta instead of Al.
}
\label{tabcr}
\begin{center}
\begin{tabular}{lcc}
             & AC off        & AC on         \\
\hline
\multicolumn{2}{l}{a) AC plastic}\\
\cline{1-1}
photons; LED & $9.5 \pm 0.9$ & $9.2 \pm 0.9$ \\
photons; HED & $33  \pm 1  $ & $32  \pm 1  $ \\
protons; LED & $271 \pm 3  $ & $14  \pm 1  $ \\
protons; HED & $316 \pm 2  $ & $9.0 \pm 0.2$ \\
\hline
\multicolumn{2}{l}{b) AC NaI}\\
\cline{1-1}
photons; LED & $7.5 \pm 1.3$ & $6.9 \pm 1.2$ \\
photons; HED & $23.3\pm 1.1$ & $21.5\pm 1.1$ \\
protons; LED & $331 \pm 3  $ & $13  \pm 1  $ \\
protons; HED & $428 \pm 2  $ & $9.9 \pm 0.3$ \\
\hline
\multicolumn{2}{l}{c) AC plastic + Ta box}\\
\cline{1-1}
photons; LED & $2.3 \pm 0.5$ & $2.2 \pm 0.5$ \\
photons; HED & $9.9 \pm 0.5$ & $9.1 \pm 0.5$ \\
protons; LED & $322 \pm 13 $ & $15  \pm 3  $ \\
protons; HED & $395 \pm 7  $ & $12  \pm 2  $ \\
\hline
\end{tabular}
\end{center}
\end{table}


The total background count rate is in the range of $2~10^{-4}$ \cntrt~in LED
and twice as much in HED.

Some ways to reduce it are proposed and under study, among them:
- to replace the plastic scintillator in AC by a crystal such as NaI, CsI or
LaBr$_3$;
- to increase the thickness of the tantalum shield in order to increase the
shielding power against photons.
However, this leads to an increase of the mass of material, and consequently of
the number of proton interactions and count rates.
The first results with a NaI AC
indicate only a slight improvement of the photon-induced background,
counterbalanced by a small increase of the proton-induced background.
A study with the higher density scintillators LaBr$_3$ is in progress.

\begin{figure}[]
\resizebox{\hsize}{!}{\includegraphics[clip=true]{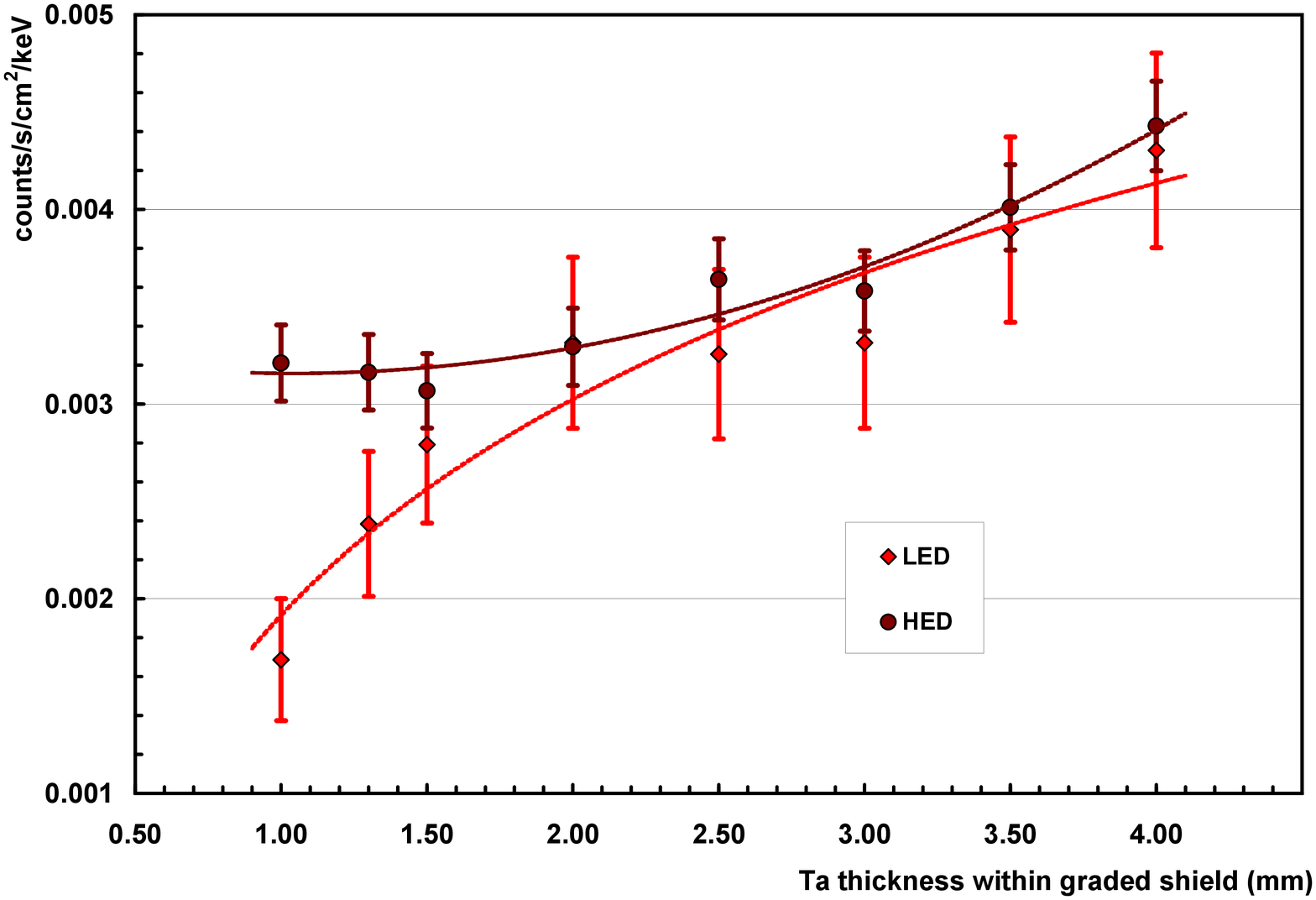}}
\caption{
\footnotesize
Variation of the count rates in detectors due to cosmic protons as function of
the tantalum thickness in graded shield.
The nominal thickness is 1.3~mm.
The solid lines are guides for the eyes.
}
\label{background3}
\end{figure}

Increasing the tantalum thickness in the graded shield to 3~mm reduces the
photon-induced part of the background on the HED by 25~\%.
However, as shown by Fig.~\ref{background3}, the thicker material layer gives
rise to a higher background due to protons, resulting in a higher total detector
background.
However, these additional events are almost entirely tagged by AC.

An alternative could be to place this supplementary tantalum shield outside of
the AC.
A preliminary study with a setup where the 2.3~mm thick aluminum of the
protective enclosure was replaced by tantalum shows that it has a large effect
on the photon-induced background:
A reduction by a factor 5 in LED and 3 in HED is observed, as shown in
table~\ref{tabcr},
to be compared with the 25~\% reduction obtained by an
increase of the tantalum thickness of the internal graded shield, while the
effect on the proton-induced background remains more or less the same.
This may indicate that the present mechanical setup suffers from indirect
leaks.
Optimization of the shielding geometry should be done to suppress these leaks.
One should note that the spacecraft, which is sitting below the detector box,
will also act as a shield.

\begin{figure}[]
\resizebox{\hsize}{!}{\includegraphics[clip=true]{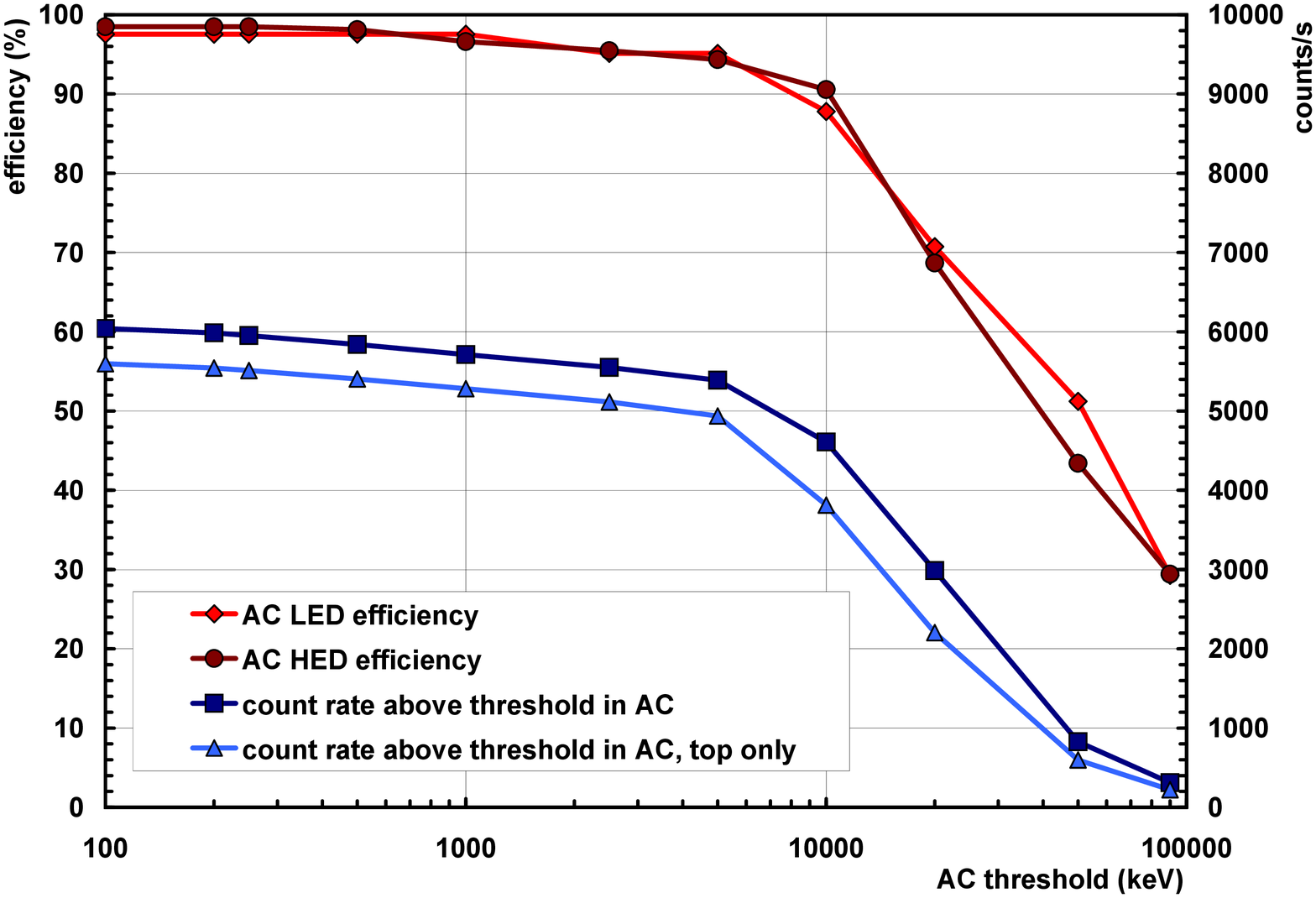}}
\caption{
\footnotesize
AC (plastic) performance (left scale) and count rates (right scale) resulting
from cosmic protons as function of the AC threshold.
}
\label{ACefficiency}
\end{figure}

As indicated in Fig.~\ref{ACefficiency}, the count rate in AC due to cosmic
protons ranges around 6~kHz, resulting in a large dead time of the LED, not
easily sustainable by this detector.
If one considers only the upper part of the AC (``top'' curve in the figure),
the count rate decreases only to $\approx$5.5~kHz.
Other schemes could however be considered to decrease rate and dead time:
different AC segmentation, optimization of the veto procedure, (for example
using correlation between LED, HED and AC) etc.

\subsection{Induced radioactivity}
\label{radioactivity}
Radioactivity induced in the spacecraft by cosmic rays leads to delayed
background in the detectors.
First results indicate that the contribution of the delayed background
to the overall background is of the order of 0.5~\%.
However, radioactivation is very dependent on materials and their
locations, and thus would be more precisely evaluated with a more detailed and
realistic mass model of the spacecraft.
%

\subsection{Optics}
\label{optics}
\begin{figure}[]
\resizebox{\hsize}{!}{\includegraphics[clip=true]{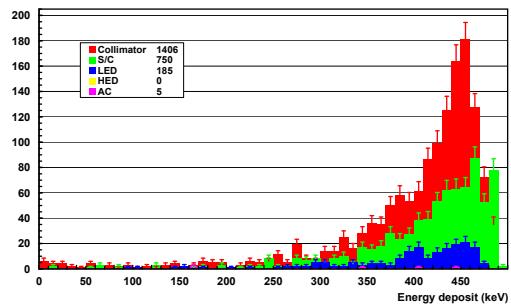}}
\caption{
\footnotesize
Energy deposit in LED, HED, AC, collimator and structure (S/C) for 500 keV
protons scattered in the mirror shells.
}
\label{mirrorEdep}
\end{figure}

A preliminary analysis of the energy deposit of 500 keV protons scattered into
the mirror shells has been performed.
About $10^6$ protons have been randomly generated over the surface of an
annular source on top of the shells, with a source half-angle of 0.5~degrees.
The detector spacecraft has been divided into collimator, structure (called S/C
in Fig.\ref{mirrorEdep}), LED, HED and AC, and energy deposits in these
different parts were recorded.
The results, shown in Fig.~\ref{mirrorEdep}, indicate that, with the generated
statistic, no protons or secondary reaches the HED.

\section{Conclusions}

Our Geant4 simulations have shown that an active anticoincidence around the
detectors is absolutely required to achieve a background level of the order
of 10$^{-4}$ \cntrt.
It has thus become apparent that the present design should be optimized toward
a more compact and above all hermetic coverage of the detectors by the AC.
However, in the same time, different schemes should be actively studied to
decrease AC count rate and LED dead time.
At this preliminary state of study, neither the induced radioactivity nor the
scattering of soft protons in the optics do seem to be a major problem
compared to the prompt background.

\begin{acknowledgements}
This work was partly supported by Bundesministerium f\"ur Wirtschaft und
Technologie through Deutsches Zentrum f\"ur Luft- und Raumfahrt e.V. (DLR)
grants FKZ 50QR0601 and FKZ 50OG0601.

\end{acknowledgements}

\bibliographystyle{aa}

\begin{thebibliography}{}

\bibitem[Agostinelli et al.(2003)]{Agos03}
Agostinelli~S. et al.,
Nucl. Instrum. and Meth. A 506 (2003) 250-303.

\bibitem[Allison et al.(2006)]{Alli06}
Allison~J. et al.,
IEEE Trans. Nucl. Sci 53 (2006) 270-278.

\bibitem[Claret(2006)]{Clar06}
Claret~A.,
``Space environment of Simbol-X'',
CEA/Dapnia/SAp internal report, 2006.

\bibitem[Gruber et al.(1999)]{Grub99}
Gruber~D.E. et al.,
\aj, 520 (1999) 124.

\bibitem[Nartallo(2002)]{Nart02}
Nartallo~R.,
``The propagation of low-energy protons through the XMM-Newton optics'',
Esa/estec/tos-ema/02-067/RN, Issue~1, revision~0;
\url{http://space-env.esa.int/Feedback.} \url{htm?document=Project
  Support/} \url{XMM Newton/tos-em-02-067.pdf}

\bibitem[O'Gallagher and Maslyar III(1976)]{Gall76}
O'Gallagher~J.J. and Maslyar~III~G.A.,
J. Geophys. Res. 81 (1976) 1319.

\bibitem[Tenzer(2006)]{Tenz06}
Tenzer~C. et al.,
Proc. SPIE 6266 (2006) 626620.

\end{thebibliography}

\end{document}